\documentclass[pdflatex,sn-mathphys-num]{sn-jnl}

\usepackage{graphicx}%
\usepackage{multirow}%
\usepackage{amsmath,amssymb,amsfonts}%
\usepackage{amsthm}%
\usepackage{mathrsfs}%
\usepackage[title]{appendix}%
\usepackage{xcolor}%
\usepackage{textcomp}%
\usepackage{manyfoot}%
\usepackage{booktabs}%
\usepackage{algorithm}%
\usepackage{algorithmicx}%
\usepackage{algpseudocode}%
\usepackage{listings}%
\usepackage{bm}%
\usepackage{braket}%
\usepackage{nicematrix}%
\usepackage{quantikz}%
\usepackage{url}%
\usepackage{hyperref}%
\usepackage{microtype}%
\usepackage{caption}
\definecolor{structure}{RGB}{0,70,127}

\theoremstyle{thmstyleone}%

\theoremstyle{thmstyletwo}%

\theoremstyle{thmstylethree}%

\begin{document}

\title[A Continuous-Variable Quantum Fourier Layer]{A Continuous-Variable Quantum Fourier Layer: Applications to Filtering and PDE Solving}

\author*[1]{\fnm{Paolo} \sur{Marcandelli}}\email{paolo.marcandelli@polimi.it}
\author[1]{\fnm{Stefano} \sur{Mariani}}\email{stefano.mariani@polimi.it}
\author[1]{\fnm{Martina} \sur{Siena}}\email{martina.siena@polimi.it}
\author[2]{\fnm{Stefano} \sur{Markidis}}\email{markidis@kth.se}

\affil*[1]{\orgdiv{Department of Civil and Environmental Engineering}, \orgname{Politecnico di Milano}, \orgaddress{\city{Milan}, \postcode{20133}, \country{Italy}}}

\affil[2]{\orgdiv{Department of Computational Science and Technology}, \orgname{KTH Royal Institute of Technology}, \orgaddress{\city{Stockholm}, \country{Sweden}}}

\abstract{Fourier representations play a central role in operator learning methods for partial differential equations and are increasingly being explored in quantum machine learning architectures. The classical fast Fourier transform (FFT), particularly in its Cooley--Tukey decomposition, exhibits a structure that naturally matches continuous-variable quantum circuits. This correspondence establishes a direct structural isomorphism between the Cooley-Tukey butterfly network and Gaussian photonic gates, enabling the FFT to be realized as a native optical computation in continuous-variable quantum computing.
Building on this observation, we introduce a continuous-variable Quantum Fourier Layer (CV--QFL) based on a bipartite Gaussian encoding and a Cooley-Tukey quantum Fourier transform, enabling exact two-dimensional spectral processing within a Gaussian photonic circuit.
We test the CV--QFL on two representative tasks: spectral low-pass filtering and Fourier-domain integration of the heat equation. In both cases, the results match the classical reference to machine precision. Beyond these examples, our method naturally extends to optical-input settings in which the signal is already available as a Gaussian optical field. In such scenarios, coherent light coupled into single-mode waveguides can be processed directly by the CV--QFL, bypassing the need for an explicit classical-to-quantum encoding stage. This enables native spectral processing of light and lays the groundwork for new approaches to quantum scientific machine learning, in particular for future neural operator architectures within the CV framework.}

\keywords{Continuous-variable quantum computing, quantum photonics, Gaussian quantum circuits, quantum Fourier transform, Fourier spectral layer, PDE solving}

\maketitle
\section{Introduction}

Fourier transforms are a fundamental tool in signal processing, imaging, and scientific computing, enabling the representation of signals and data in the spectral domain and supporting efficient operations such as filtering and spectral methods for the numerical solution of partial differential equations. For discrete signals, the introduction of the fast Fourier transform (FFT), most prominently through the Cooley--Tukey (CT) algorithm~\cite{d3ea2d52-5ab2-3128-8b80-efb85267295d}, reduced the computational cost of the discrete Fourier transform from $\mathcal{O}(N^2)$ to $\mathcal{O}(N \log N)$ for signals of size $N$ by recursively decomposing the transform into smaller operations. Since then, FFT methods have become a cornerstone of modern signal processing and scientific computing \cite{duhamel1990fft,oppenheim1999dsp,brigham1988fft}. 

Beyond digital implementations, Fourier transforms can also be realized directly in optical systems, where interference and propagation naturally perform spectral transformations~\cite{goodman2005fourier,psaltis1988optical}. Moreover, the butterfly structure of the CT decomposition maps naturally onto optical operations such as beam splitting and phase shifting, motivating photonic implementations of the FFT in integrated optical circuits~\cite{nejadriahi2017opticalfft,ahmed2020photonicfft,Barak}. While these approaches have mainly been explored for classical optical signal processing, their connection with continuous-variable (CV) quantum computing remains comparatively less developed. Establishing such a connection is especially relevant given the central role played by the quantum Fourier transform (QFT) in quantum algorithms and, more recently, in spectral-based quantum machine learning (QML) models. In this work, we bridge this gap by identifying a structural correspondence between the CT algorithm and its realization in CV photonic quantum circuits, where the elementary operations of the butterfly network map naturally onto Gaussian transformations implemented through beam splitters and phase shifts. This perspective provides the basis for more advanced photonic architectures for spectral processing, with potential applications in QML. In particular, the proposed Continuous-Variable Quantum Fourier layer (CV-QFL) is designed to process classical data defined on two-dimensional domains, making it naturally suited to problems involving structured fields and mappings between functions, as in neural operator frameworks~\cite{JMLR:v24:21-1524,li2020fno}. However, this framework can also be extended to intrinsically quantum settings and is therefore not limited to classical problems, as we discuss in Sec.~\ref{sec:discussion}.

\noindent \textbf{Quantum machine learning and the problem of encoding.} A central challenge in QML is how to encode classical data into quantum states. To construct a CV--QFL, one must first determine an encoding strategy that is compatible with the target computational task. Whether the goal is signal filtering, the spectral solution of partial differential equations, or operator learning, the chosen encoding determines the class of Fourier transformations that can be realized within the circuit and consequently the spectral information accessible to the computation~\cite{math12213318,Schuld_2021}. 

In standard qubit-based frameworks, several encoding strategies have been proposed, including basis encoding, amplitude encoding, time-evolution encoding, and Hamiltonian encoding~\cite{Nielsen_Chuang_2010,munikote2024comparing,Kok_Lovett_2010}. These approaches typically map classical data onto a single computational register, naturally leading to one-dimensional QFT implementations~\cite{Mart_n_L_pez_2012,PhysRevLett.86.1889,Griffiths_1996}. Such constructions admit efficient decompositions in terms of Hadamard and controlled-phase gates, but remain intrinsically tied to a single-register spectral structure. This limitation becomes particularly relevant when addressing more complex tasks involving two- or three-dimensional fields, such as operator learning for partial differential equations.

Unary encoding has recently emerged as an alternative paradigm~\cite{Landman_2022,Ramos_Calderer_2021,johri2020nearestcentroidclassificationtrapped,kerenidis2022quantummachinelearningsubspace, Xiao2025quantumdeeponet}, where the restriction to a fixed-excitation subspace enables the exact realization of orthogonal transformations and enhanced amplitude control. This makes it especially suitable for operator-learning architectures~\cite{JMLR:v24:21-1524,johri2021nearest,lu2021deeponet}. Within this setting, Fourier operators \cite{li2020fno} have been implemented in one-dimensional quantum architectures~\cite{jain2024qfno} and extended to multidimensional domains through tensorized multi-register constructions~\cite{marcandelli2025phqfno,pfeffer2023multidimensionalquantumfouriertransformation}, often leveraging CT factorizations decomposed into phase shifters and reconfigurable beam splitter (RBS) gates that preserve excitation number~\cite{d3ea2d52-5ab2-3128-8b80-efb85267295d}. While naturally suited to linear-optical platforms, these constructions generally require nontrivial compilation overhead on standard qubit hardware~\cite{PhysRevLett.125.120504}. 

These observations motivate the investigation of QFT implementations formulated directly in terms of optical primitives within the CV setting. In CV quantum computing, information is encoded in bosonic modes with continuous spectra rather than in discrete two-level systems, leading to an infinite-dimensional Hilbert space with enhanced representational capacity and a natural compatibility with structured spectral transformations \cite{Lloyd_1999,Weedbrook_2012,lau_2017,Schuld_2019}. Several CV encoding strategies have been proposed for quantum machine learning, ranging from feature-map embeddings to Gaussian-state encodings \cite{lau_2017,killoran2019continuous,rath2025continuousvariablequantumencodingtechniques}. However, in most existing approaches, classical data are embedded either as single vectors or as local pointwise features \cite{Markidis_2022,panichi2025quantumphysicsinformedneural}. While such strategies are suitable for regression or classification, they do not represent the full discretized field as a globally structured object at the quantum-state level. This becomes a serious limitation in operator learning and scientific machine learning, where the data naturally arise as multidimensional fields defined over spatial grids. In such settings, the lack of a structured global encoding prevents the direct application of multidimensional spectral transformations within the quantum circuit. Even when global encodings are considered, they remain predominantly one-dimensional or restricted to discrete-variable constructions \cite{rath2025continuousvariablequantumencodingtechniques,Gottesman_2001}, thereby constraining the class of multidimensional Fourier operators that can be realized.

\noindent \textbf{Cooley-Tukey Fourier transforms in CV quantum computing.} To address the problem of implementing two-dimensional Fourier transforms in a CV quantum framework, we introduce a bipartite Gaussian encoding scheme in which the input matrix is embedded directly into the cross-correlation blocks of a multimode covariance matrix. By exploiting inter-register entanglement as a computational resource, the input field is represented at the state level as a structured two-dimensional object, preserving its separable spatial organization already at the encoding stage. This induces a tensor-product organization of the bipartite register and naturally enables a genuinely two-dimensional QFT within a Gaussian CV architecture.

Within this setting, the elementary operations of the CT decomposition map directly onto Gaussian photonic transformations implemented through beam splitters and phase shifts acting on the two registers. By exploiting this separable structure, the Fourier transform can be performed independently along each register, thereby realizing a two-dimensional QFT of the input state within a CV photonic circuit. The resulting transform exhibits a computational complexity scaling as \(O(m \log m + n \log n)\) for an \(m \times n\) domain, compared with the \(O(mn \log(mn))\) cost of a classical two-dimensional FFT applied to a flattened grid. To the best of our knowledge, this is the first CV--QFL that explicitly integrates CV quantum circuits with the CT Fourier-transform framework.

More broadly, the proposed construction establishes a new paradigm for spectral processing in CV quantum computing. Combined with the bipartite Gaussian encoding, it provides a natural route for mapping two-dimensional inputs into the Fourier domain through Gaussian states of bosonic modes, with potential applications in quantum machine learning, operator learning, signal filtering, and quantum-inspired solvers for partial differential equations. It also provides a foundation for photonic architectures capable of performing spectral processing directly within a CV quantum computing framework.

%\subsection{Main contributions}
To summarize, the main contributions of this work are as follows:
\begin{itemize}
    \item \textbf{Bipartite Gaussian matrix encoding.}
    We introduce a CV encoding scheme that embeds a matrix 
    $D \in \mathbb{R}^{m \times n}$ exactly into the cross-covariance block of a bipartite Gaussian state, yielding a structured and physically implementable representation of two-dimensional data in CV systems.
    
    \item \textbf{CV--QFL.}
    We design a CV--QFL acting independently on two registers, enabling an exact two-dimensional quantum Fourier transform of the encoded matrix within a Gaussian photonic architecture.

    \item \textbf{Spectral signal filtering and PDE solving.}
    We validate the proposed Fourier layer on two representative tasks: denoising of a two-dimensional signal corrupted by Gaussian noise, and spectral solution of the two-dimensional heat equation directly in the Fourier domain.
\end{itemize}
\section{Methods}\label{sec:methods}

This section presents the methodological foundations of the proposed framework. We first introduce the bipartite encoding procedure used to map two-dimensional classical data onto the covariance matrix of a multimode Gaussian state. Building on this representation, we then describe the implementation of the two-dimensional Fourier transform through a CV--QFL, which constitutes the core spectral layer of the proposed approach. A theoretical overview of the basic concepts of CV quantum circuits, Gaussian states and covariance matrices is provided in Appendix~\ref{app:thbackground}.

\subsection{Bipartite Entanglement-Based Encoding}\label{sec:encoding}
We consider a bipartite CV architecture composed of two registers, $r_1$ and $r_2$, containing $n$ and $m$ bosonic modes, respectively, for a total of $n+m$ modes. \\
To encode classical correlations between two sets of modes, we employ the two-mode squeezing (TMS) operator:  
\begin{equation}
\hat S_2(r,\phi)
=
\exp\!\left[
r\left(
e^{i\phi}\hat a_i\hat a_j
-
e^{-i\phi}\hat a_i^\dagger\hat a_j^\dagger
\right)
\right]
\end{equation}
acting between a mode $i$ in register $r_1$ and a mode $j$ in register $r_2$.
For $\phi=0$, its action on the quadrature operators reads
\begin{equation}
\hat{\bm r}_{ij}
\;\longmapsto\;
S_{\mathrm{TMS}}(r)\,\hat{\bm r}_{ij},\quad \text{with}\quad S_{\mathrm{TMS}}(r)=\begin{pmatrix}
\cosh r & \sinh r & 0 & 0 \\
\sinh r & \cosh r & 0 & 0 \\
0 & 0 & \cosh r & -\sinh r \\
0 & 0 & -\sinh r & \cosh r
\end{pmatrix}
\end{equation}
Starting from the vacuum state ($\hbar=2$), where $\langle \hat x_i^2\rangle=\langle \hat x_j^2\rangle=1$ and $\langle \hat x_i \hat x_j\rangle=0$, one can show that $\langle \hat x_i' \hat x_j' \rangle=\sinh(2r).$
A naive entry-wise encoding of a data matrix 
$D \in \mathbb{R}^{m \times n}$ would assign independent squeezing parameters
\begin{equation}
r_{ij} = \tfrac12 \operatorname{arcsinh}(\lambda D_{ij}),
\end{equation}
so that $\langle \hat x_i^{r_1} \hat x_j^{r_2} \rangle = \lambda D_{ij}$. 
Here, $\lambda > 0$ is a global scaling factor introduced to control the squeezing amplitudes and keep the required two-mode squeezing operations within physically reasonable energy bounds. 
However, such an entry-wise construction is not implementable in practice: two-mode squeezing gates acting on shared modes do not commute, and successive applications alter both local variances and previously established inter-register correlations. As a result, it is not possible to independently encode all entries $D_{ij}$ in this manner, since later operations inevitably distort the covariance structure generated by earlier ones.

To overcome this limitation, the encoding is performed in a decoupled basis obtained via the Singular Value Decomposition (SVD) of the input matrix,
\begin{equation}\label{eq:svd}
D = U \Sigma V^\top,
\end{equation}
where $\Sigma=\mathrm{diag}(\sigma_1,\dots,\sigma_r)$. In this basis, each singular value $\sigma_k$ corresponds to an independent mode pair, allowing the assignment
\begin{equation}
r_k=\tfrac12\operatorname{arcsinh}(\lambda\sigma_k)
\end{equation}
without interference between different correlations. In this way, one can embed the diagonal matrix $\Sigma$ directly into the inter-register cross-covariance block of the Gaussian state without inter register interference. 
Then, the resulting covariance matrix acquires the block structure

\[
\bm{\sigma}_{\mathrm{TMS}} =
\begin{pNiceMatrix}[
  cell-space-limits = 5pt
]
            C_{r_1}
         & \fbox{\(\Sigma\)}
         & 0
         & 0  \\[4pt]
   \Sigma^\top
         & C_{r_2}
         & 0
         & 0 \\[4pt]
 0
         & 0
         & C_{r_1}
         & \fbox{\(-\Sigma\)} \\[4pt]
  0
         & 0
         & -\Sigma^\top
         & C_{r_2}
\end{pNiceMatrix}
\]

where the subscripts $r_1$ and $r_2$ denote the quadratures associated with registers $1$ and $2$, respectively, $C_{r_1} = \mathrm{diag}(\cosh 2r_1, \ldots, \cosh 2r_k, 1, \ldots, 1) \in \mathbb{R}^{m \times m}$
and, analogously, $C_{r_2} \in \mathbb{R}^{n \times n}$ collect the local thermal occupation terms of each mode, while $\Sigma \in \mathbb{R}^{m \times n}$ is the diagonal matrix of singular values, with $\Sigma_{ii} = \sinh(2r_i) = \sigma_i$. We notice that the off-diagonal block $\sigma_{x_{r_1} x_{r_2}}$ contains the inter-register $xx$ correlations and coincides, in the rotated SVD basis, with the encoded matrix $\Sigma$. \\
As described by the universality theorem \cite{clements2017optimaldesignuniversalmultiport}, both factors $U$
and $V$ of Eq. \ref{eq:svd} are physically realizable as passive interferometers on the
respective registers. By the application of this theorem, it will be possible to reconstruct the original input matrix $D$.\\
When $U$ acts on $r_1$ and $V$ acts simultaneously on $r_2$, one can show by following Appendix \ref{subsec:unitary} about Gaussian unitary evolution, that the full symplectic transformation on the joint quadrature vector
$\hat{\boldsymbol{r}} = (x_{r_1}, x_{r_2}, p_{r_1}, p_{r_2})^\top$ is
\begin{equation}\label{eq:simplectic}
    S_{\mathrm{tot}} =
    \begin{pmatrix}
        \mathrm{Re}(U) & 0              & -\mathrm{Im}(U) & 0              \\
        0              & \mathrm{Re}(V) &  0              & -\mathrm{Im}(V)\\
        \mathrm{Im}(U) & 0              &  \mathrm{Re}(U) & 0              \\
        0              & \mathrm{Im}(V) &  0              &  \mathrm{Re}(V)
    \end{pmatrix},
\end{equation}
and that the full covariance matrix transforms as
$\bm{\sigma} \mapsto S_{\mathrm{tot}}\,\bm{\sigma}\,S_{\mathrm{tot}}^\top$.
Extracting the cross-block
$\bm{\sigma}_{x_{r_1} x_{r_2}}$ from this expression,
one obtains
\begin{equation}
    \bm{\sigma}_{x_{r_1} x_{r_2}}
    \;\longmapsto\;
    \mathrm{Re}(U)\,\Sigma\,\mathrm{Re}(V)^\top
    - \mathrm{Im}(U)\,\Sigma\,\mathrm{Im}(V)^\top.
\end{equation}
For a real data matrix $D \in \mathbb{R}^{m\times n}$ with SVD
$D = U\Sigma V^\top$, the unitary factors $U$ and $V$ are real orthogonal
($\mathrm{Im}(U) = \mathrm{Im}(V) = 0$), and the expression above reduces to
\begin{equation}\label{eq:encoded}
    \boxed{
    \bm{\sigma}_{x_{r_1} x_{r_2}} \;=\; U\,\Sigma\,V^\top \;=\; D.
    }
\end{equation}
The data matrix $D$ is therefore encoded exactly
in the $x$-quadrature cross-block of the bipartite covariance matrix,
with $U$ acting from the left via register $r_1$ and $V$ from the right via
register $r_2$, see Fig. \ref{fig:cv_fourier_layer} for the complete CV quantum circuit representation.
The cross-block $\bm{\sigma}_{p_{r_1} p_{r_2}} = -D$ acquires the
opposite sign, while all $x$-$p$ cross-correlations vanish identically. \\

For additional details on the theoretical aspects of the encoding procedure such as complexity and bipartite entanglement, and on its Strawberry Fields implementation, see Appendix~\ref{app:encodingdetails}.

\subsection{Cooley-Tukey as Continuous-Variable Quantum circuit}\label{sec:cooleytukey}
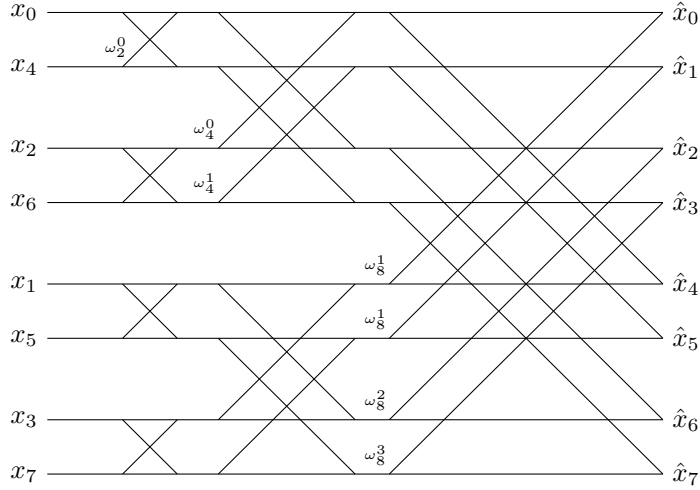
\begin{figure}[htb]
\centering
\begin{tikzpicture}[x=0.9cm, y=0.9cm]

    % Group 1
    \node[left] at (0,0) {$x_0$};
    \node[right] at (9,0) {$\hat{x}_0$};
    \draw (0,0) -- (9,0);
    \node[left] at (0,-0.8) {$x_4$};
    \node[right] at (9,-0.8) {$\hat{x}_1$};
    \draw (0,-0.8) -- (9,-0.8);

    % Group 2
    \node[left] at (0,-2) {$x_2$};
    \node[right] at (9,-2) {$\hat{x}_2$};
    \node[above] at (4.8,-6) {\tiny$\omega_8^2$};
    \node[above] at (4.8,-6.8) {\tiny$\omega_8^3$};
    \draw (0,-2) -- (9,-2);
    \node[left] at (0,-2.8) {$x_6$};
    \node[right] at (9,-2.8) {$\hat{x}_3$};
    \draw (0,-2.8) -- (9,-2.8);

    % Group 3
    \node[left] at (0,-4) {$x_1$};
    \node[above] at (1,-0.8) {\tiny$\omega_2^0$};
    \node[above] at (2.3,-2) {\tiny$\omega_4^0$};
    \node[above] at (2.3,-2.8) {\tiny$\omega_4^1$};
    \node[above] at (4.8,-4) {\tiny$\omega_8^1$};
    \node[above] at (4.8,-4.8) {\tiny$\omega_8^1$};
    \node[right] at (9,-4) {$\hat{x}_4$};
    \draw (0,-4) -- (9,-4);
    \node[left] at (0,-4.8) {$x_5$};
    \node[right] at (9,-4.8) {$\hat{x}_5$};
    \draw (0,-4.8) -- (9,-4.8);

    % Group 4
    \node[left] at (0,-6) {$x_3$};
    \node[right] at (9,-6) {$\hat{x}_6$};
    \draw (0,-6) -- (9,-6);
    \node[left] at (0,-6.8) {$x_7$};
    \node[right] at (9,-6.8) {$\hat{x}_7$};
    \draw (0,-6.8) -- (9,-6.8);

    \def\crosssize{0.4}
    \coordinate (C1) at (1.5,-0.4);
    \draw ($(C1)+(-\crosssize,-\crosssize)$) -- ($(C1)+(\crosssize,\crosssize)$);
    \draw ($(C1)+(-\crosssize,\crosssize)$) -- ($(C1)+(\crosssize,-\crosssize)$);
    \coordinate (C2) at (1.5,-2.4);
    \draw ($(C2)+(-\crosssize,-\crosssize)$) -- ($(C2)+(\crosssize,\crosssize)$);
    \draw ($(C2)+(-\crosssize,\crosssize)$) -- ($(C2)+(\crosssize,-\crosssize)$);
    \coordinate (C3) at (1.5,-4.4);
    \draw ($(C3)+(-\crosssize,-\crosssize)$) -- ($(C3)+(\crosssize,\crosssize)$);
    \draw ($(C3)+(-\crosssize,\crosssize)$) -- ($(C3)+(\crosssize,-\crosssize)$);
    \coordinate (C4) at (1.5,-6.4);
    \draw ($(C4)+(-\crosssize,-\crosssize)$) -- ($(C4)+(\crosssize,\crosssize)$);
    \draw ($(C4)+(-\crosssize,\crosssize)$) -- ($(C4)+(\crosssize,-\crosssize)$);

    \def\crosssizes{1}
    \coordinate (C5) at (3.5,-1);
    \draw ($(C5)+(-\crosssizes,-\crosssizes)$) -- ($(C5)+(\crosssizes,\crosssizes)$);
    \draw ($(C5)+(-\crosssizes,\crosssizes)$) -- ($(C5)+(\crosssizes,-\crosssizes)$);
    \coordinate (C6) at (3.5,-1.8);
    \draw ($(C6)+(-\crosssizes,-\crosssizes)$) -- ($(C6)+(\crosssizes,\crosssizes)$);
    \draw ($(C6)+(-\crosssizes,\crosssizes)$) -- ($(C6)+(\crosssizes,-\crosssizes)$);
    \coordinate (C7) at (3.5,-5);
    \draw ($(C7)+(-\crosssizes,-\crosssizes)$) -- ($(C7)+(\crosssizes,\crosssizes)$);
    \draw ($(C7)+(-\crosssizes,\crosssizes)$) -- ($(C7)+(\crosssizes,-\crosssizes)$);
    \coordinate (C8) at (3.5,-5.8);
    \draw ($(C8)+(-\crosssizes,-\crosssizes)$) -- ($(C8)+(\crosssizes,\crosssizes)$);
    \draw ($(C8)+(-\crosssizes,\crosssizes)$) -- ($(C8)+(\crosssizes,-\crosssizes)$);

    \def\crosssizess{2}
    \coordinate (C9) at (7,-2);
    \draw ($(C9)+(-\crosssizess,-\crosssizess)$) -- ($(C9)+(\crosssizess,\crosssizess)$);
    \draw ($(C9)+(-\crosssizess,\crosssizess)$) -- ($(C9)+(\crosssizess,-\crosssizess)$);
    \coordinate (C10) at (7,-2.8);
    \draw ($(C10)+(-\crosssizess,-\crosssizess)$) -- ($(C10)+(\crosssizess,\crosssizess)$);
    \draw ($(C10)+(-\crosssizess,\crosssizess)$) -- ($(C10)+(\crosssizess,-\crosssizess)$);
    \coordinate (C11) at (7,-4);
    \draw ($(C11)+(-\crosssizess,-\crosssizess)$) -- ($(C11)+(\crosssizess,\crosssizess)$);
    \draw ($(C11)+(-\crosssizess,\crosssizess)$) -- ($(C11)+(\crosssizess,-\crosssizess)$);
    \coordinate (C12) at (7,-4.8);
    \draw ($(C12)+(-\crosssizess,-\crosssizess)$) -- ($(C12)+(\crosssizess,\crosssizess)$);
    \draw ($(C12)+(-\crosssizess,\crosssizess)$) -- ($(C12)+(\crosssizess,-\crosssizess)$);

\end{tikzpicture}
\caption{Butterfly diagram of the CT algorithm for the FFT of
the bit-reversed input $x\in\mathbb{R}^8$.
Each cross is a radix-2 butterfly with twiddle factor
$\omega_n^k = e^{-i2\pi k/n}$.}
\label{CooleyTukey}
\end{figure}

In this section, we present the core idea of our work, namely the structural isomorphism between the CT algorithm and the set of optical operations available in CV quantum circuits. \\
We begin by reviewing the implementation of the CT algorithm to implement the FFT in the one-dimensional setting, before extending the construction to higher-dimensional domains. We then show how its computational structure maps naturally onto CV circuit operations, which forms the central idea underlying this work. To have more information about the classical DFT before approaching the CT algorithm, see Appendix \ref{app:dft}.\\
To Fourier transform a one-dimensional vector $x\in\mathbb{R}^N$, the
procedure begins by permuting its indices according to a bit-reversal scheme.
Throughout this work we adopt the direct-transform sign convention
$\omega_n^k = e^{-i\frac{2\pi k}{n}}$.
The core computational step, with $N=8$ as an example, is the butterfly-shaped cross operation,
see Fig.~\ref{CooleyTukey}.

In this diagram, each cross
\begin{center}
\begin{tikzpicture}
    \draw (0,0) -- (3,0);
    \node[above] at (0.8,-0.8) {$\omega_n^k$};
    \node[left]  at (0,0) {$a$};
    \node[right] at (3,0) {$a+\omega_n^k b$};
    \draw (0,-0.8) -- (3,-0.8);
    \node[left]  at (0,-0.8) {$b$};
    \node[right] at (3,-0.8) {$a-\omega_n^k b$};
    \def\crosssize{0.4}
    \coordinate (C1) at (1.5,-0.4);
    \draw ($(C1)+(-\crosssize,-\crosssize)$) -- ($(C1)+(\crosssize,\crosssize)$);
    \draw ($(C1)+(-\crosssize,\crosssize)$)  -- ($(C1)+(\crosssize,-\crosssize)$);
\end{tikzpicture}
\end{center}
is applied iteratively to pairs of elements and maps $[a,b]$ as follows:
\begin{equation}\label{eq:butterfly}
    \begin{pmatrix} 1 & \omega_n^k \\ 1 & -\omega_n^k \end{pmatrix}
    \begin{pmatrix} a \\ b \end{pmatrix}
    =
    \begin{pmatrix} a + \omega_n^k b \\ a - \omega_n^k b \end{pmatrix},
    \qquad \omega_n^k = e^{-i\frac{2\pi k}{n}},
\end{equation}
where $n=2,4,8$ is the number of inputs at each stage and
$k\in\{0,1,\ldots,\tfrac{n}{2}-1\}$.

\medskip
In the CV circuit, each butterfly cross is replaced by a
sequence of two optical gates acting on a pair of modes $(q_i,q_j)$, as shown in Fig.~\ref{fig:cv_butterfly}:

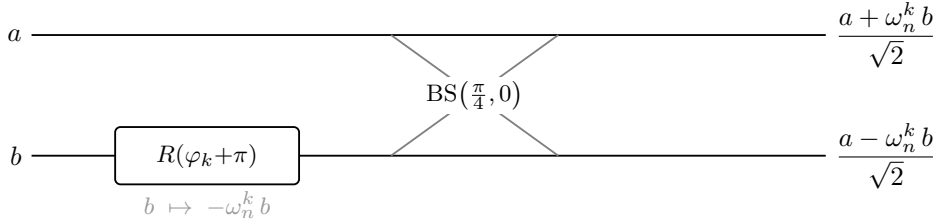
\begin{figure}[htb]
\centering
\begin{tikzpicture}[x=1.1cm, y=1.0cm, line width=0.7pt]

  \def\ya{0}
  \def\yb{-1.6}

  \draw (0,\ya) -- (9.5,\ya);
  \draw (0,\yb) -- (9.5,\yb);

  \node[left] at (0,\ya) {$a$};
  \node[left] at (0,\yb) {$b$};

  % Rgate
  \filldraw[fill=white, draw=black, rounded corners=2pt]
    (1.0, \yb-0.38) rectangle (3.2, \yb+0.38);
  \node[font=\small] at (2.1,\yb) {$R(\varphi_k {+} \pi)$};
  \node[font=\scriptsize, text=gray!80] at (2.1,\yb-0.65)
    {$b \;\mapsto\; {-}\omega_n^k\, b$};

  % BSgate (X cross only)
  \draw[black!50, line width=0.7pt] (4.3,\ya) -- (6.3,\yb);
  \draw[black!50, line width=0.7pt] (4.3,\yb) -- (6.3,\ya);
  \node[font=\small, fill=white, inner sep=1.5pt]
    at (5.3,{(\ya+\yb)/2})
    {$\mathrm{BS}\!\left(\tfrac{\pi}{4},0\right)$};

  % Output labels
  \node[right] at (9.5,\ya) {$\dfrac{a + \omega_n^k\, b}{\sqrt{2}}$};
  \node[right] at (9.5,\yb) {$\dfrac{a - \omega_n^k\, b}{\sqrt{2}}$};

\end{tikzpicture}
\caption{CV implementation of a single CT butterfly.
  The phase gate $R(\varphi_k{+}\pi)$ maps $b\mapsto -\omega_n^k b$;
  the beam splitter $\mathrm{BS}(\pi/4,0)$ then produces
  $(a\pm\omega_n^k b)/\sqrt{2}$, identical to the classical
  output~\eqref{eq:butterfly} up to the $1/\sqrt{2}$
  ortho-normalisation factor.}
\label{fig:cv_butterfly}
\end{figure}

The Rgate maps mode $q_j$ as
$b \mapsto e^{i(\varphi_k+\pi)}\,b = -\omega_n^k\, b$,
where $\varphi_k = -2\pi k/n$, so that $e^{i\varphi_k} = \omega_n^k$.
The BSgate, whose symplectic action is
$a'=(a-b)/\sqrt{2}$, $b'=(a+b)/\sqrt{2}$, then gives
\[
  \frac{1}{\sqrt{2}}
  \begin{pmatrix} 1 & \omega_n^k \\ 1 & -\omega_n^k \end{pmatrix}
  \begin{pmatrix} a \\ b \end{pmatrix}
  =
  \frac{1}{\sqrt{2}}
  \begin{pmatrix} a + \omega_n^k b \\ a - \omega_n^k b \end{pmatrix}.
\]
In this way, we reconstruct the transformation we find in the classical
butterfly~\eqref{eq:butterfly}, realised entirely through operations
that are native to CV photonic circuits.
Notably, the $1/\sqrt{2}$ normalisation per stage arises automatically
from the unitarity of the beam splitter.
After these operations, in both circuits the output is the desired Fourier transformed vector.\\

We are now ready to extend the one-dimensional CT transformation to the two-dimensional setting. As shown in Eq.~\eqref{eq:2dfft_matrix} of the Appendix \ref{app:dft}, the 2D discrete Fourier transform admits a separable structure, and can therefore be implemented through independent one-dimensional transforms applied along each dimension. The classical CT radix-2
algorithm~\cite{d3ea2d52-5ab2-3128-8b80-efb85267295d}
exploits this separability by first applying a 1D FFT of
length $n$ to each of the $m$ rows of $M$, at a cost of
$\mathcal{O}(n \log n)$ per row, and subsequently applying
a 1D FFT of length $m$ to each of the $n$ columns of the
intermediate result, at a cost of $\mathcal{O}(m \log m)$
per column. The total arithmetic cost is therefore

\begin{equation}\label{eq:fft}
    \text{gate count:}\quad m \cdot \mathcal{O}(n \log n)
    \;+\;
    n \cdot \mathcal{O}(m \log m)
    \;=\;
    \mathcal{O}(mn \log(mn)).
\end{equation}

\subsection{Continuous-Variable Quantum Fourier Layer}\label{sec:cv-qfl}

We now show how the CV--QFL can be physically implemented in CV quantum computing 
by applying independent single-register Fourier transforms to a bipartite Gaussian state encoding the matrix $D$ in its cross-correlation block.
To this end, we now apply Fourier matrices $F_m$ to register $r_1$ and $F_n$ to register $r_2$ simultaneously.
Then, in the same way of Eq. \eqref{eq:simplectic} and Eq. \ref{eq:2dfft_matrix} of Appendix \ref{app:dft}, the global symplectic matrix takes the explicit block form
\begin{equation} \label{eq:xR1transform2d}
    S_{\mathrm{tot}} =
    \begin{pmatrix}
        \mathrm{Re}(F_m) & 0               & -\mathrm{Im}(F_m) & 0                \\[4pt]
        0               & \mathrm{Re}(F_n) & 0                & -\mathrm{Im}(F_n) \\[4pt]
        \mathrm{Im}(F_m) & 0               &  \mathrm{Re}(F_m) & 0                \\[4pt]
        0               & \mathrm{Im}(F_n) & 0                &  \mathrm{Re}(F_n)
    \end{pmatrix},
\end{equation}
so that the full quadrature transformation is compactly expressed as
\begin{equation}\label{eq:xR2transform2d}
    \hat{\boldsymbol r} \;\longmapsto\; S_{\mathrm{tot}}\,\hat{\boldsymbol r},
    \qquad
    \bm{\sigma} \;\longmapsto\; S_{\mathrm{tot}}\,\bm{\sigma}\,S_{\mathrm{tot}}^\top.
\end{equation}

Substituting~\eqref{eq:xR1transform2d} and~\eqref{eq:xR2transform2d} into
$\bm{\sigma}_{x_{r_1}x_{r_2}}'{}_{ij}
= \langle \hat{x}_i^{\prime r_1}\,\hat{x}_j^{\prime r_2}\rangle$
and using the four inter-register second moments of the encoded state,
\begin{equation}
    \langle \hat{x}_k^{r_1}\hat{x}_l^{r_2}\rangle = D_{kl},
    \quad
    \langle \hat{p}_k^{r_1}\hat{p}_l^{r_2}\rangle = -D_{kl},
    \quad
    \langle \hat{x}_k^{r_1}\hat{p}_l^{r_2}\rangle
    = \langle \hat{p}_k^{r_1}\hat{x}_l^{r_2}\rangle = 0,
\end{equation}
one obtains
\begin{align}
    \bm{\sigma}_{x_{r_1}x_{r_2}}'{}_{ij}
    &= \sum_{k,l}
        \mathrm{Re}(F_m)_{ik}\,\mathrm{Re}(F_n)_{jl}
        \underbrace{\langle\hat{x}_k^{r_1}\hat{x}_l^{r_2}\rangle}_{D_{kl}}
      - \sum_{k,l}
        \mathrm{Re}(F_m)_{ik}\,\mathrm{Im}(F_n)_{jl}
        \underbrace{\langle\hat{x}_k^{r_1}\hat{p}_l^{r_2}\rangle}_{0}
        \nonumber\\
    &\quad
      - \sum_{k,l}
        \mathrm{Im}(F_m)_{ik}\,\mathrm{Re}(F_n)_{jl}
        \underbrace{\langle\hat{p}_k^{r_1}\hat{x}_l^{r_2}\rangle}_{0}
      + \sum_{k,l}
        \mathrm{Im}(F_m)_{ik}\,\mathrm{Im}(F_n)_{jl}
        \underbrace{\langle\hat{p}_k^{r_1}\hat{p}_l^{r_2}\rangle}_{-D_{kl}},
\end{align}
which simplifies to
\begin{equation} \label{eq:block1}
    \bm{\sigma}_{x_{r_1}x_{r_2}}'
    = \mathrm{Re}(F_m)\,D\,\mathrm{Re}(F_n)^\top
    - \mathrm{Im}(F_m)\,D\,\mathrm{Im}(F_n)^\top
    = \mathrm{Re}\!\left(F_m\,D\,F_n^\top\right).
\end{equation}
Note that, unlike the one-dimensional case, the $pp$ block
$\langle\hat{p}^{r_1}\hat{p}^{r_2}\rangle = -D$ now contributes
a non-trivial term because both registers are transformed. Then, substituting~\eqref{eq:xR1transform2d} and~\eqref{eq:xR2transform2d} into $\bm{\sigma}_{x_{r_1}p_{r_2}}'{}_{ij}
= \langle \hat{x}_i^{\prime r_1}\,\hat{p}_j^{\prime r_2}\rangle$ one similarly has:
\begin{equation} \label{eq:block2}
    \bm{\sigma}_{x_{r_1}p_{r_2}}'{}_{ij} = \mathrm{Re}(F_m)\,D\,\mathrm{Im}(F_n)^\top
     + \mathrm{Im}(F_m)\,D\,\mathrm{Re}(F_n)^\top
    \;=\; \mathrm{Im}\!\left(F_m\,D\,F_n^\top\right).
\end{equation}

The two blocks \eqref{eq:block1} and \eqref{eq:block2} combine into a
single complex quantity:
\begin{align}
    \bm{\sigma}_{x_{r_1}x_{r_2}}' + i\,\bm{\sigma}_{x_{r_1}p_{r_2}}'
    &= \mathrm{Re}\!\left(F_m D F_n^\top\right)
     + i\,\mathrm{Im}\!\left(F_m D F_n^\top\right)
     = F_m\,D\,F_n^\top,
\end{align}
so that
\begin{equation}
    \boxed{
    \hat{D} \;=\; F_m\,D\,F_n^\top
    \;=\; \bm{\sigma}_{x_{r_1}x_{r_2}}' + i\,\bm{\sigma}_{x_{r_1}p_{r_2}}'.
    }
\end{equation}
This is precisely the two-dimensional DFT of $D$
(see Eq.~\eqref{eq:2dfft_matrix}).
The result can be read directly from the covariance matrix:

\[
\bm{\sigma}' =
\begin{pNiceMatrix}[
  cell-space-limits = 5pt,
  code-for-first-row = \color{black}\scriptstyle,
  code-for-first-col = \color{black}\scriptstyle
]
\sigma'_{x_{r_1} x_{r_1}}
        & \boxed{\mathrm{Re}(\hat{D})}
        & \sigma'_{x_{r_1} p_{r_1}}
        & \boxed{\mathrm{Im}(\hat{D})} \\[4pt]
\boxed{\mathrm{Re}(\hat{D})^{\!\top}}
        & \sigma'_{x_{r_2} x_{r_2}}
        & \boxed{\mathrm{Im}(\hat{D})^{\!\top}}
        & \sigma'_{x_{r_2} p_{r_2}} \\[4pt]
\sigma'_{p_{r_1} x_{r_1}}
        & \boxed{\mathrm{Im}(\hat{D})}
        & \sigma'_{p_{r_1} p_{r_1}}
        & \boxed{-\mathrm{Re}(\hat{D})} \\[4pt]
\boxed{\mathrm{Im}(\hat{D})^{\!\top}}
        & \sigma'_{p_{r_2} x_{r_2}}
        & \boxed{-\mathrm{Re}(\hat{D})^{\!\top}}
        & \sigma'_{p_{r_2} p_{r_2}}
\end{pNiceMatrix}
\]
where the boxed entries are entirely determined by $\hat{D} = F_m D F_n^\top$,
and the unboxed diagonal blocks describe local single-register variances.
The two-dimensional DFT of $D$ is therefore obtained directly from the
second moments of the bipartite Gaussian state, without any additional
processing: it suffices to read the cross-block
$\bm{\sigma}_{x_{r_1}x_{r_2}}'$ and $\bm{\sigma}_{x_{r_1}p_{r_2}}'$
of the covariance matrix $\bm{\sigma}'$ after the CT circuits
have been applied to both registers, and combine them as
\begin{equation}\label{eq:QFT}
    \hat{D} \;=\; \bm{\sigma}_{x_{r_1}x_{r_2}}' \;+\; i\,\bm{\sigma}_{x_{r_1}p_{r_2}}',
\end{equation}
where the real part encodes the cosine components of the transform
and the imaginary part the sine components.
Both blocks are entries of the physical covariance matrix and are
directly accessible via homodyne measurements on the two registers. \\

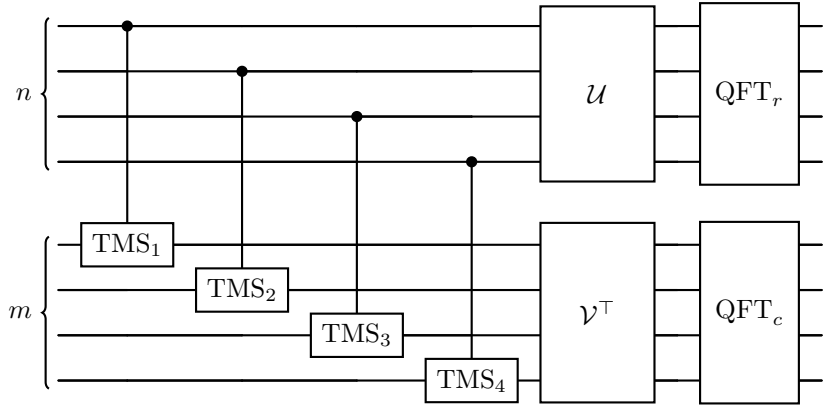
\begin{figure}[h]
\centering
\begin{quantikz}[row sep={0.6cm,between origins}, column sep=0.3cm]
%
% ── Row register ─────────────────────────────────────────────────
\lstick[4]{$n$}
  & \ctrl{4} & \qw & \qw & \qw
  & \gate[4][1.5cm]{\mathcal{U}} & \qw
  & \gate[4][1.3cm]{\mathrm{QFT}_r} & \qw \\
  & \qw & \ctrl{4} & \qw & \qw
  & & \qw & & \qw \\
  & \qw & \qw & \ctrl{4} & \qw
  & & \qw & & \qw \\
  & \qw & \qw & \qw & \ctrl{4}
  & & \qw & & \qw \\[0.5cm]
%
% ── Column register ───────────────────────────────────────────────
\lstick[4]{$m$}
  & \gate{\mathrm{TMS}_1} & \qw & \qw & \qw
  & \gate[4][1.5cm]{\mathcal{V}^\top} & \qw
  & \gate[4][1.3cm]{\mathrm{QFT}_c} & \qw \\
  & \qw & \gate{\mathrm{TMS}_2} & \qw & \qw
  & & \qw & & \qw \\
  & \qw & \qw & \gate{\mathrm{TMS}_3} & \qw
  & & \qw & & \qw \\
  & \qw & \qw & \qw & \gate{\mathrm{TMS}_4}
  & & \qw & & \qw \\
\end{quantikz}
\caption{%
  CV--QFL circuit for an $m \times n$ input matrix
  on $m+n$ optical modes.
  \textbf{Encoding}: TMS gates inject the singular values of $M$;
  interferometers $\mathcal{U}$ and $\mathcal{V}^\top$ reconstruct
  the input matrix via SVD.
  \textbf{Fourier Layer}: the CT QFT is applied
  independently to both registers, yielding the 2D optical
  Fourier transform.
}
\label{fig:cv_fourier_layer}
\end{figure}

A complete graphical representation of the CV--QFL circuit is provided in Fig.~\ref{fig:cv_fourier_layer}. The figure highlights the three main stages of the architecture: singular-value encoding via two-mode squeezing gates, the $U$ and $V$ interferometers used to reconstruct the input matrix $D$, and the CT QFT decomposition applied independently to the two registers.

The two-dimensional CT QFT application results in
a circuit of depth $\mathcal{O}(\log m)$ on $r_1$
and a circuit of depth $\mathcal{O}(\log n)$ on $r_2$.
Since the two registers are disjoint, the two circuits
run in parallel, giving
\begin{equation}\label{eq:qft}
    \text{gate count: }
    \mathcal{O}(m \log m + n \log n),
    \qquad
    \text{depth: }
    \mathcal{O}(\log(mn)).
\end{equation}
A direct comparison between Eqs.~\eqref{eq:fft} and \eqref{eq:qft} shows that the bipartite CV implementation achieves gate complexity $\mathcal{O}(m\log m + n\log n)$, as opposed to the classical $\mathcal{O}(mn\log(mn))$, thereby reducing the circuit resources required to implement the two-dimensional Fourier transform.

\section{Results}
\label{sec:applications}

To illustrate the practical utility of the CV--QFL pipeline, we present two
concrete applications: low-pass spectral filtering and numerical solution of a
diffusive partial differential equation.
In both cases the computation is carried out entirely inside the photonic
circuit using StrawberryFields.

% ─────────────────────────────────────────────────────────────────────────────
\subsection{Spectral Low-Pass Filtering}
\label{sec:filtering}

We consider a $64\times64$ test image composed of three known low-frequency components,
\begin{equation}
  s(r,c)
  = \cos\!\left(\tfrac{2\pi\cdot3\,r}{64}\right)\cos\!\left(\tfrac{2\pi\cdot3\,c}{64}\right)
  + \tfrac{1}{2}\cos\!\left(\tfrac{2\pi\cdot7\,r}{64}\right)
  + \tfrac{1}{2}\cos\!\left(\tfrac{2\pi\cdot5\,c}{64}\right),
\end{equation}
corrupted by additive white Gaussian noise with unit standard deviation, yielding an input SNR (signal-to-noise ratio) of $-3.0\,\mathrm{dB}$.

\begin{figure}[h]
  \centering
  \includegraphics[width=\linewidth]{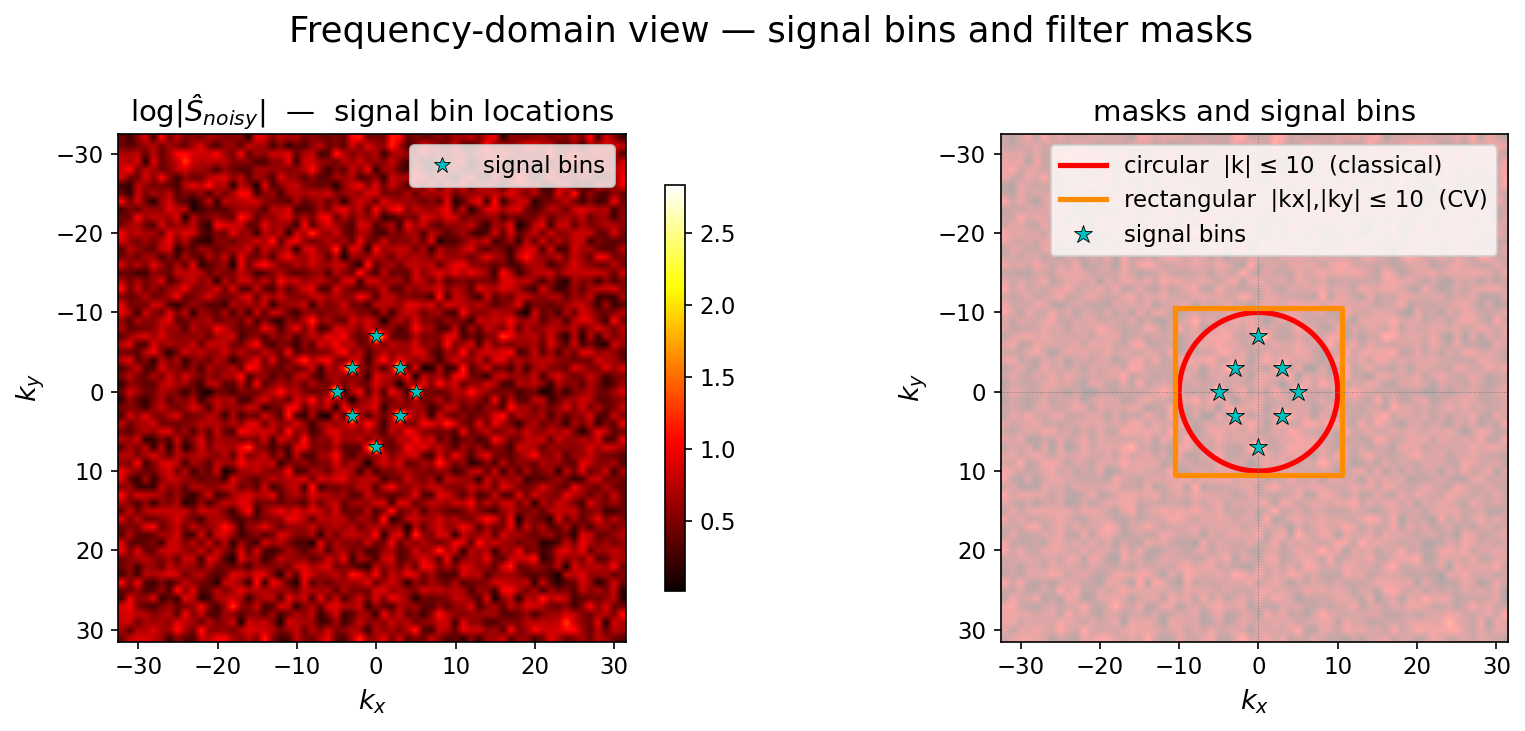}
\caption{Comparison of the spectral masks used in the filtering experiment: a circular mask with $|\mathbf{k}| \leq 10$ and a separable rectangular mask with $|k_x| \leq 10$ and $|k_y| \leq 10$. The signal frequency components lie within both masks.}
  \label{fig:masks}
\end{figure}

Because each cosine term generates discrete Fourier peaks (equivalently, delta-like contributions on specific frequency bins), the spectrum of $s$ is exactly sparse. In centered frequency notation, its nonzero components are located at
\begin{equation}
  \{(\pm3,\pm3),\;(\pm7,0),\;(0,\pm5)\},
  \label{eq:signal_bins}
\end{equation}
as shown in Fig. \ref{fig:masks}.
Thus, the signal energy is concentrated on only 8 Fourier bins associated with the prescribed low-frequency components, while the added noise is broadband and distributed across the full spectrum. Therefore, by preserving the few low-frequency bins containing the signal and suppressing the remaining noise-dominated modes, the inverse Fourier transform reconstructs the original image with high fidelity and improved SNR ratio.\\

To implement low-pass filtering in the CV--QFL, we use the \texttt{LossChannel} operation available in Strawberry Fields. Physically, \texttt{LossChannel}$(T)$ models optical attenuation by mixing the signal mode with a vacuum mode through a beam splitter of transmissivity $T\in[0,1]$. In the Gaussian formalism, its action on the covariance matrix $\bm\sigma$ and mean vector $\mu$ is
\begin{equation}
  \bm\sigma \longrightarrow T\,\bm\sigma + (1-T)\,\mathbf{I},
  \qquad
  \mu \longrightarrow \sqrt{T}\,\mu.
\end{equation}
At $T=1$ the transformation is the identity, while at $T=0$ the mode is replaced by vacuum and all correlations with the rest of the system are removed.

A key property of \texttt{LossChannel} is that it is phase-insensitive: both quadratures of a mode are attenuated by the same factor $\sqrt{T}$. As a result, when \texttt{LossChannel} is applied independently to row and column modes, each complex Fourier coefficient
\begin{equation}
  z[i,j] = \sigma_{xx}[i,m{+}j] + i\,\sigma_{xp}[i,m{+}j]
\end{equation}
is simply rescaled as
\begin{equation}\label{eq:losschannel}
  z[i,j] \longrightarrow \sqrt{T_iT_j}\,z[i,j].
\end{equation}
This is essential, since it suppresses selected Fourier amplitudes without altering their phase.

To realize a low-pass filter, we choose binary transmissivities $T_i,T_j\in\{0,1\}$ so that only modes within the desired frequency range are preserved, while higher-frequency modes are removed. This implements a separable rectangular spectral mask in Fourier space. After the inverse QFT, the filtered signal is reconstructed directly from the output covariance matrix, without any intermediate digital manipulation of the spectrum.
\paragraph{Spectral Filtering Results: }We compare the proposed CV--QFL low-pass filtering pipeline against a classical Fourier baseline on a $64\times64$ signal corrupted by Gaussian noise. In the classical pipeline, the signal is first mapped to the frequency domain through a two-dimensional FFT, filtered with a circular low-pass mask of cutoff $\|\mathbf{k}\|\leq 10$, and finally reconstructed by the inverse FFT. In the CV--QFL implementation, the signal is first transformed through the optical QFT, then filtered by applying \texttt{LossChannel} to suppress modes outside the selected frequency range, and finally reconstructed through the inverse QFT, obtained as its adjoint operation. Both the circular mask $|\mathbf{k}|\leq 10$ used in the classical pipeline and the separable rectangular mask $|k_x|\leq 10$, $|k_y|\leq 10$ used in the CV implementation contain the spectral bins as shown in Fig. \ref{fig:masks}.\\

Table~\ref{tab:filtering} summarizes the quantitative comparison. The CV circuit improves the signal-to-noise ratio by $+9.4\,\mathrm{dB}$, close to the $+10.7\,\mathrm{dB}$ achieved by the classical pipeline. The small gap is explained by the larger rectangular mask, which retains slightly more noise power than the circular classical one. Importantly, when the CV--QFL output is compared against a classical reference computed with the same rectangular mask, the maximum pointwise error is $3.8\times10^{-15}$, i.e. at machine precision. This confirms that the photonic circuit reproduces the intended spectral filtering exactly.

\begin{table}[h]
  \centering
  \begin{tabular}{lcc}
    \toprule
    & \textbf{Classical} & \textbf{CV-QFL} \\
    \midrule
    Mask shape            & circular           & rectangular (separable) \\
    Bins retained         & 317 / 4096 (7.7\%) & 441 / 4096 (10.8\%) \\
    SNR in (dB)           & $-3.0$             & $-3.0$ \\
    SNR out (dB)          & $7.7$              & $6.4$ \\
    SNR improvement (dB)  & $+10.7$            & $+9.4$ \\
    Error vs same-mask ref.  & ---             & $3.8\times10^{-15}$ \\
    \bottomrule
  \end{tabular}
  \caption{Spectral filtering comparison on a $64\times64$ grid.}
  \label{tab:filtering}
\end{table}

\begin{figure}[h]
  \centering
  \includegraphics[width=\linewidth]{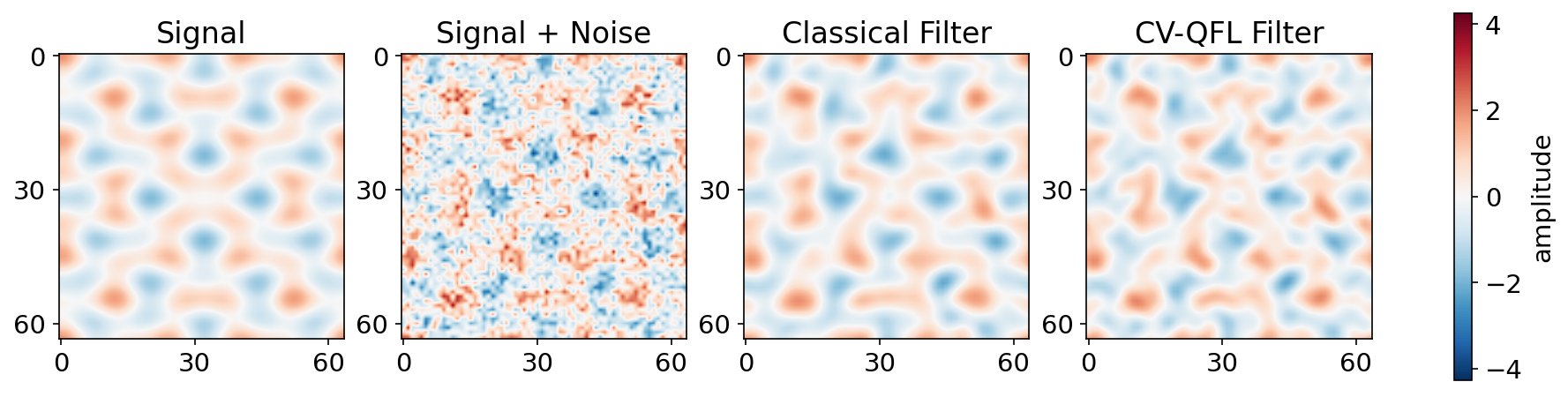}
  \caption{Spectral low-pass filtering on a $64\times64$ image.
    \emph{Left to right}: clean signal; noisy input ($\mathrm{SNR}=-3.0\,\mathrm{dB}$);
    classical output (circular mask, $+10.7\,\mathrm{dB}$);
    CV--QFL output (rectangular separable mask, $+9.4\,\mathrm{dB}$).
    The photonic circuit implements the rectangular spectral mask to
    machine-precision accuracy.}
  \label{fig:filtering_spatial}
\end{figure}

Figure~\ref{fig:filtering_spatial} shows the clean signal, the noisy input, and the outputs of the classical and CV filtering pipelines. In both cases, the low-frequency structure is successfully preserved while most of the broadband noise is removed.

% ─────────────────────────────────────────────────────────────────────────────
\subsection{Two-dimensional Heat Equation}
\label{sec:heat}

We consider the two-dimensional heat equation
\begin{equation}
  \frac{\partial u}{\partial t} = \alpha \nabla^2 u,
  \qquad \alpha = 0.05.
\end{equation}
A convenient way to solve this equation is in the Fourier domain, where the Laplacian acts diagonally and the dynamics decouple mode by mode. As a result, each Fourier coefficient evolves independently according to an exponential damping law,
\begin{equation}
  \hat{u}(\mathbf{k},t+\Delta t)
  = e^{-\alpha |\mathbf{k}|^2 \Delta t}\,\hat{u}(\mathbf{k},t).
\end{equation}
This defines the exact spectral propagator
\begin{equation}
  G(\mathbf{k},\Delta t)=e^{-\alpha |\mathbf{k}|^2 \Delta t},
\end{equation}
which attenuates high-frequency modes more strongly than low-frequency ones. The corresponding solution strategy therefore follows the same structure as the filtering task: the field is transformed to the Fourier domain, evolved through the mode-wise propagator, and then reconstructed by the inverse Fourier transform.\\
In this example, we consider a $32\times32$ spatial grid, with an initial condition given by the superposition of two localized Gaussian peaks. Because such a profile contains substantial mid-to-high spatial frequency content, it provides a convenient test for Fourier-domain diffusion. The system is evolved for four time steps of size $\Delta t=0.2$ up to a final time $t=0.8$, during which the peaks progressively broaden and decrease in amplitude, as expected from diffusive heat dynamics, see Fig. \ref{fig:filtering_spatial}.\\
After the forward QFT, applying \texttt{LossChannel}$(T_i)$ to row mode $i$ and
\texttt{LossChannel}$(T_j)$ to column mode $j$ scales the cross-covariance as shown in Eq. \ref{eq:losschannel}. Then, choosing
$T_i = e^{-2\alpha k_{r,i}^2\,\Delta t}$ and
$T_j = e^{-2\alpha k_{c,j}^2\,\Delta t}$
gives
\begin{equation}
  \sqrt{T_i\,T_j}
  = e^{-\alpha(k_{r,i}^2+k_{c,j}^2)\Delta t}
  = G(\mathbf{k},\Delta t),
\end{equation}
i.e.\ the exact heat-equation propagator, where subscripts $r$ and $c$ stand for rows and column registers.
The only difference from the low-pass filter is the transmissivity profile
programmed into \texttt{LossChannel}: binary $\{0,1\}$ for filtering,
$e^{-2\alpha k^2\Delta t}$ for diffusion.

\paragraph{Heat Equation Results}
At every time step the CV--QFL output agrees with the classical pseudospectral
solver to floating-point precision as shown in table \ref{tab:heat_error}.\\
\begin{table}[h!]
\centering
\begin{tabular}{cc}
  \toprule
  Time $t$ & $\max \lvert u_{\mathrm{CV}} - u_{\mathrm{classical}} \rvert$ \\
  \midrule
  0.0 & $0.0\times10^{0}$ \\
  0.2 & $4.4\times10^{-16}$ \\
  0.4 & $5.0\times10^{-16}$ \\
  0.6 & $5.6\times10^{-16}$ \\
  0.8 & $6.4\times10^{-16}$ \\
  \bottomrule
\end{tabular}
\captionsetup{width=\textwidth,font=small, justification=justified, singlelinecheck=false}
\caption{Maximum absolute error between the classical pseudospectral
solution and the solution computed with the CV--QFL at different
time instants for the heat equation. The error remains at the level of
floating-point precision over the entire temporal evolution.}
\label{tab:heat_error}
\end{table}

\begin{figure}[h]
  \centering
  \includegraphics[width=\linewidth]{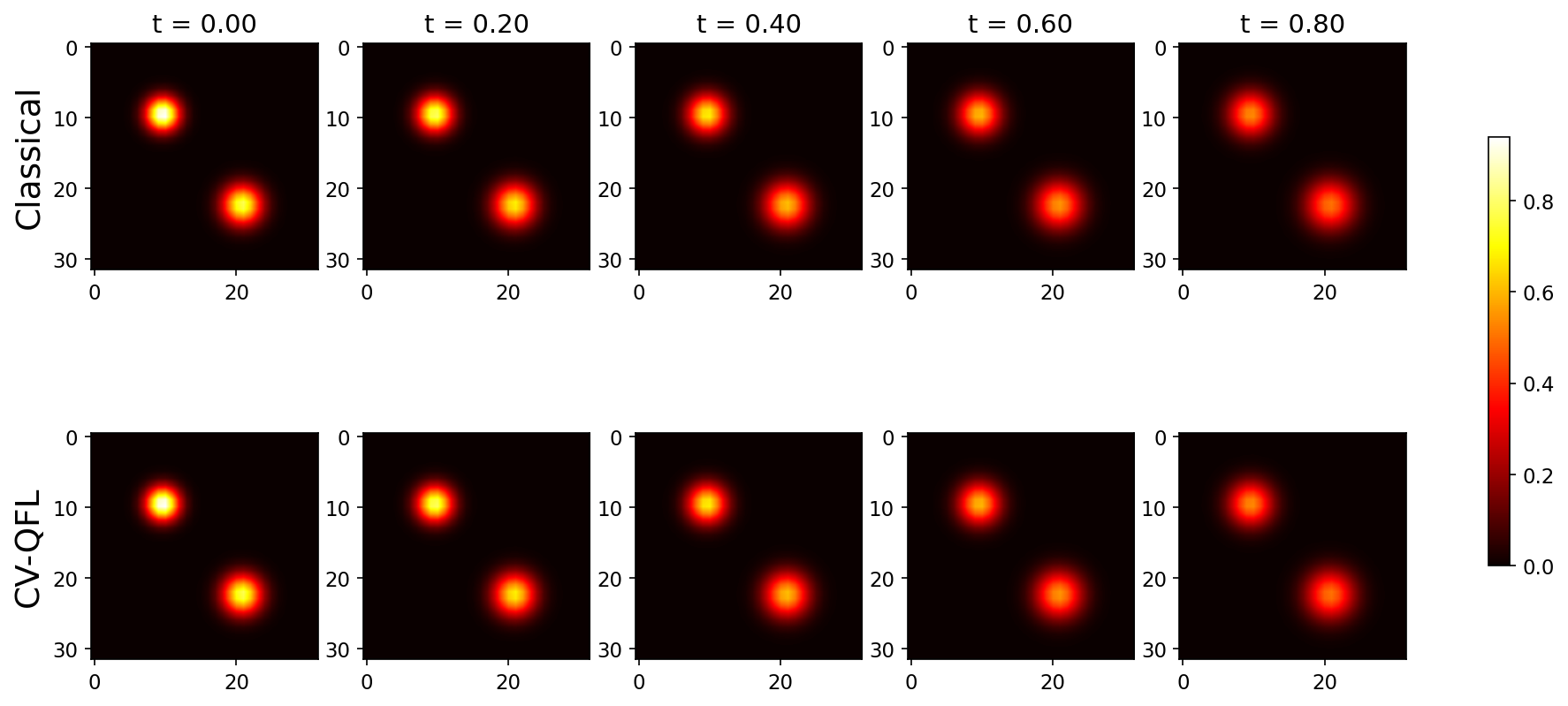}
  \caption{Heat equation integration on a $32\times32$ grid
    ($\alpha=0.05$, $\Delta t=0.2$, four steps).
    \emph{Top row}: classical pseudospectral solver (NumPy).
    \emph{Bottom row}: CV--QFL photonic circuit.
    The two rows are identical to floating-point precision at every snapshot.}
  \label{fig:heat_spatial}
\end{figure}

The two Gaussian blobs spread and merge under diffusion;
Figure~\ref{fig:heat_spatial} shows that the classical and CV--QFL
snapshots are visually indistinguishable at all time steps, confirming that \texttt{LossChannel}$(T_k = e^{-2\alpha k^2\Delta t})$ implements the exact heat-equation propagator inside the optical circuit.

\section{Discussion}\label{sec:discussion}
A particularly promising direction for this framework concerns signals that are natively optical, such as coherent fields generated by interferometers, wavefront sensors, or free-space optical systems. In these settings, the input is already carried by optical modes and can be processed directly within the photonic circuit, bypassing the need for explicit classical-to-quantum encoding steps. This makes the proposed CV-QFL uniquely suited for hardware-native spectral processing and for quantum machine learning scenarios where data is intrinsically optical.

Furthermore, the utility of the CV-QFL extends to natively quantum problems where information is directly encoded within the covariance matrix of multi-mode states. In such cases, classical architectures face a significant computational bottleneck, as the classical simulation of high-dimensional Gaussian state transformations becomes increasingly resource-intensive. By enabling the direct manipulation of these quantum correlations, the CV-QFL architecture serves as a versatile tool capable of addressing both classical signal processing and quantum-specific tasks that lie beyond the efficient reach of classical digital computers. This provides a robust foundation for future research into fully quantum-native frameworks where the CV-QFL can demonstrate a natural quantum advantage.

A second important perspective is the connection with operator learning. Unlike standard CV encodings, which typically represent data as vectors of local amplitudes or single-mode features, the present construction embeds an entire two-dimensional field into the cross-covariance structure of a bipartite Gaussian state. As a result, the CV-QFL acts on a structured field rather than on isolated scalar inputs. This is especially relevant for scientific machine learning models such as Fourier Neural Operators \cite{li2020fno}. In this sense, the proposed framework may provide a natural CV photonic foundation for hardware-native spectral operator learning. 

However, achieving a definitive quantum advantage for purely classical problems remains challenging. Even highly efficient Quantum Neural Operators are fundamentally constrained by the classical-to-quantum data encoding bottleneck, which requires significant computational overhead and large-scale hardware management that is difficult to realize in near-term devices. Consequently, outperforming classical architectures on purely classical datasets remains very challenging. Conversely, we believe the true potential of our architecture lies in its application as a neural model for inherently quantum-mechanical Hamiltonian problems. In such contexts, classical neural operators struggle due to the intractability of simulating large-scale quantum dynamics. This opens a novel research avenue: integrating the CV--QFL into a quantum neural operator within a deep learning framework dedicated to quantum simulations.

At the same time, the present framework has important limitations. First, the proposed cross-covariance encoding is intrinsically two-dimensional, so extending it to higher-dimensional or spatio-temporal fields will require new register structures. Second, the current architecture is entirely Gaussian and therefore efficiently simulable classically, which prevents any claim of computational quantum advantage. Extending the framework with non-Gaussian resources could overcome this limitation and lead to more expressive quantum models, although at the price of substantially greater experimental complexity. Finally, the present implementation relies on separable mode-by-mode spectral processing; exploring richer forms of inter-register coupling may enable more general classes of spectral operators in future developments.
\section{Conclusions}
\label{sec:conclusions}
In this work, we introduced a CV--QFL based on a structural correspondence between the CT decomposition and CV quantum circuits, where beam splitters and phase shifters naturally implement the elementary operations of the butterfly network. Building on this observation, we developed a bipartite Gaussian encoding that represents a two-dimensional input matrix in the cross-covariance structure of a multimode photonic state, enabling an exact two-dimensional quantum Fourier transform within a CV framework.

In particular, the input matrix is encoded through its singular value decomposition: the singular values are embedded via TMS gates, while the original matrix is reconstructed by independently applying the interferometers associated with \(U\) and \(V\) to the two registers. On this basis, we establish a structural correspondence between the classical CT butterfly network and the native gate set of CV quantum computing, leading to a two-dimensional QFT acting on the off-diagonal block \(\sigma_{xx}\) of the covariance matrix. Implemented entirely within a CV circuit, this CV-QFL constitutes the core contribution of the paper and provides a native spectral primitive for CV photonic platforms.

Furthermore, we showed that the proposed construction leads to a reduced gate complexity for the two-dimensional Fourier transform. In particular, the bipartite CV implementation scales as \(\mathcal{O}(m\log m + n\log n)\), in contrast to the classical \(\mathcal{O}(mn\log(mn))\), resulting in a more resource-efficient circuit for an \(m \times n\) transform.

From the experimental side, we validated the CV--QFL on two representative tasks. First, in spectral low-pass filtering of a noisy two-dimensional signal, our proposed spectral layer reproduced the target spectral mask with machine-precision accuracy and achieved signal reconstruction comparable to the classical reference. Second, in the spectral integration of the two-dimensional heat equation, the circuit reproduced the exact Fourier-domain propagator and matched the classical pseudospectral solution at machine precision across all time steps.

More broadly, the proposed framework provides a photonic route to spectral processing in CV quantum computing, with potential relevance for quantum machine learning and Fourier-based PDE solvers. While the current Gaussian implementation remains classically simulable, it establishes a foundational building block for natively quantum neural operators. Future work will focus on extending this framework to higher-dimensional domains and integrating non-Gaussian resources to address intrinsically quantum-mechanical problems, such as many-body Hamiltonian dynamics, where the direct manipulation of bosonic states can offer a clear path toward a natural quantum advantage.
% Paste here the content of Conclusion.tex

\section*{Acknowledgements}

Financial support from ICSC -- ``National Research Centre in High Performance Computing, Big Data and Quantum Computing'', funded by the European Union -- NextGenerationEU, is gratefully acknowledged.

\clearpage
\appendix

\appendix
\section{Theoretical Background}\label{app:thbackground}

We review the fundamental theoretical background necessary for understanding Sec. \ref{sec:methods} of the paper. We begin by reviewing the basic concepts of CV circuits, including Gaussian states and covariance matrices, before turning to unitary state evolution.
\subsection{Gaussian States and the Covariance Matrix}
CV quantum  systems are described in terms of bosonic modes, each corresponding to a quantum harmonic oscillator. For an $n$-mode system, we introduce the vector of canonical quadrature operators
\[
\hat{\bm r} = (\hat x_1,\dots,\hat x_n,\hat p_1,\dots,\hat p_n)^\top,
\]
where the quadratures are defined as
\begin{equation}\label{eq:quadratures}
\hat x_k = \frac{\hat a_k + \hat a_k^\dagger}{\sqrt{2}}
\qquad \text{and} \qquad
\hat p_k = \frac{\hat a_k - \hat a_k^\dagger}{i\sqrt{2}},
\end{equation}
and satisfy the canonical commutation relations
\[
[\hat x_k,\hat p_j] = i\,\delta_{kj}.
\]
Just as in classical physics, the state of the system can be described
in terms of its behaviour in phase space \cite{Kok_Lovett_2010}. The essentials of this behaviour are captured by
the Wigner function which was originally proposed by Wigner in his 1932 paper \cite{PhysRev.40.749}.  There is a special class of states that is important for optical quantum information processing with continuous variables: the so-called Gaussian states, whose Wigner function is Gaussian in phase space and are fully characterized by their first and second statistical moments, namely the displacement vector $\bm\mu = \langle \hat{\bm r} \rangle$ and the covariance matrix
\[
\sigma_{ij} = \tfrac{1}{2}\langle\{\hat r_i-\mu_i,\hat r_j-\mu_j\}\rangle,
\]
where $\{\cdot,\cdot\}$ denotes the anticommutator.
The covariance matrix is a $2n\times 2n$, real and symmetric matrix which must satisfy the uncertainty principle \cite{PhysRevA.49.1567}:
\begin{equation}\label{ineq}
\bm \sigma + \tfrac{i}{2}\,\Omega \ge 0,
\qquad
\Omega = \bigoplus_{k=1}^{n}
\begin{pmatrix}
0 & 1 \\
-1 & 0
\end{pmatrix},
\end{equation}

where $\Omega$ is the symplectic form. Examples of Gaussian states are the vacuum, coherent states, and squeezed coherent states. For more information see \cite{Braunstein_2005, Weedbrook_2012, Kok_Lovett_2010}.

\subsection{Gaussian Unitary evolution}\label{subsec:unitary}
Gaussian unitary operations correspond to affine symplectic transformations in phase space that preserve the canonical commutation relations and map Gaussian states into Gaussian states.
By the universality theorem for linear-optical networks
\cite{clements2017optimaldesignuniversalmultiport,
svozil1996experimentalrealizationdiscreteoperator},
any unitary matrix $U \in \mathrm{U}(m)$ can be exactly decomposed into a
sequence of beamsplitters and phase shifters acting on $m$ bosonic modes.
A passive interferometer implementing $U \in \mathrm{U}(m)$ acts linearly on
the annihilation operators as:
\begin{equation}
    \hat{a}_{k} \;\longmapsto\; \sum_{j=1}^{m} U_{kj}\,\hat{a}_{j}.
\end{equation}
Since the annihilation operators are the fundamental degrees of freedom
of the electromagnetic field, all passive optical transformations act
directly on $\hat{a}$ and $\hat{a}^\dagger$. The transformation of
quadratures is a derived consequence and, by using the definitions in Eq. \ref{eq:quadratures} one can show that
\begin{equation}
    \hat{\boldsymbol{r}}
    \;\longmapsto\;
    S_U\,\hat{\boldsymbol{r}}
    \qquad \text{with} \qquad 
    S_U =
    \begin{pmatrix}
        \mathrm{Re}(U) & -\mathrm{Im}(U) \\[2pt]
        \mathrm{Im}(U) &  \mathrm{Re}(U)
    \end{pmatrix}
    \in \mathrm{Sp}(2m,\mathbb{R}),
\end{equation}
and
\begin{equation}\label{eq:unevolution}
    \bm{\sigma} \;\longmapsto\; S_U\,\bm{\sigma}\,S_U^\top.
\end{equation}
In our work, we use the transformation in Eq. \ref{eq:unevolution} in Sec. \ref{sec:encoding} and in Sec. \ref{sec:cv-qfl} to construct and derive the mathematical description of the CV-QFL.
\section{CV--QFL: theoretical details and code implementation}
\label{app:encodingdetails}

This appendix collects complementary theoretical and implementation details related to the bipartite encoding scheme introduced in Sec.~\ref{sec:encoding}. We begin by discussing its entanglement structure and then analyse the optical circuit complexity of the encoding procedure. Next, we briefly review the discrete Fourier transform (DFT), which provides the necessary background for Secs.~\ref{sec:cooleytukey} and \ref{sec:cv-qfl}, where the CV--QFL is developed. Finally, we describe the Strawberry Fields implementation used in our numerical experiments.

\subsection{Bipartite Entanglement} 
While the two-mode squeezed vacuum is naturally described in the quadrature basis, it admits an equivalent representation in the Fock basis  which is very useful to understand the role of entanglement in our encoding procedure. This representation makes the bipartite entanglement structure explicit, as the state admits the Schmidt decomposition~\cite{Braunstein_2005}
\begin{equation}
\hat S_2(r)\ket{00}
=
\sqrt{1-\lambda}
\sum_{n=0}^{\infty}
\lambda^{n/2}\ket{n}\ket{n},
\qquad
\lambda=\tanh^2 r,
\end{equation}
where the Fock states $\{\ket{n}\}$ form the Schmidt basis. In this form, the entanglement can be quantified by the von Neumann entropy of the reduced density operator~\cite{van_Enk_1999},
\begin{equation}
E(r)
=
\cosh^2 r\,\log(\cosh^2 r)
-
\sinh^2 r\,\log(\sinh^2 r).
\end{equation}
In our encoding scheme, each singular value $\sigma_k$ determines a squeezing parameter $r_k$ and therefore a corresponding contribution to the total entanglement between the two registers. The entanglement spectrum of the resulting bipartite Gaussian state is thus directly determined by the singular value spectrum of $\Sigma$, establishing a one-to-one correspondence between spectral magnitude and quantum correlations. We notice that Gaussian operations applied independently on the two registers, such as the passive interferometers implementing $U$ and $V$, see Sec. \ref{sec:encoding}, correspond to local symplectic transformations and therefore leave the inter-register entanglement invariant \cite{Nielsen_Chuang_2010, Braunstein_2005, Weedbrook_2012}. 
As a result, the entanglement spectrum is entirely determined by the squeezing parameters $r_k$ associated with the singular values. 

\subsection{Circuit Complexity of the Encoding}

The encoding circuit acts on $m+n$ bosonic modes, partitioned into two registers $r_1$ ($m$ modes) and $r_2$ ($n$ modes). The construction consists of a layer of two-mode squeezing gates followed by two passive interferometers implementing the unitary factors of the SVD.

First, the $k=\min(m,n)$ two-mode squeezing (TMS) gates act on disjoint mode pairs and are therefore mutually independent. They can be applied in a single parallel layer, contributing $k$ Gaussian entangling gates with circuit depth equal to one.

Then, the unitary matrices $U\in\mathrm{U}(m)$ and $V\in\mathrm{U}(n)$ are realised through the Clements rectangular decomposition~\cite{clements2017optimaldesignuniversalmultiport}, which implements an $\ell$-mode unitary using exactly $\frac{\ell(\ell-1)}{2}$ beam-splitter--phase-shifter pairs arranged in $\ell$ columns of $\lfloor \ell/2 \rfloor$ parallel gates. Since $U$ and $V$ act on disjoint registers, the two interferometers can be executed simultaneously. Their combined cost is therefore
\[
\frac{m(m-1)}{2}+\frac{n(n-1)}{2}
= \mathcal{O}(m^2+n^2),
\qquad
\text{with depth } \mathcal{O}(\max(m,n)).
\]

The total encoding circuit is therefore dominated by the interferometric stages, yielding an overall optical gate count of $\mathcal{O}(m^2+n^2)$ and a circuit depth of $\mathcal{O}(\max(m,n))$ on $m+n$ modes. However, this circuit-level complexity should be distinguished from the cost of the full input-loading pipeline. In particular, the factorisation $D=U\Sigma V^\top$ must be computed classically in advance, with complexity $\mathcal{O}(mn\min(m,n))$, which constitutes a significant preprocessing overhead and a potential computational bottleneck for large inputs. Although this step does not contribute to the optical circuit depth once $U$, $V$, and $\Sigma$ are fixed, it remains an essential part of the overall workflow.

This observation also clarifies the scope of the proposed encoding: its main value is not to provide an immediate computational speedup for arbitrary classical data loading, but rather to identify a structured CV optical representation in which Fourier-domain processing can be implemented natively. In scenarios where the input is already available in optical form, or is generated directly by an optical front-end, this classical preprocessing may be reduced or avoided altogether, making the framework more naturally suited to native photonic data rather than generic classically stored matrices.

\subsection{Discrete Fourier Transform}\label{app:dft}
Now, we introduce the classical DFT in both one and two dimensions. We start with the definition of the DFT of a vector
$\mathbf{v} \in \mathbb{C}^l$:
\begin{equation}
\hat{v}_k
=
\frac{1}{\sqrt{l}}
\sum_{j=0}^{l-1}
v_j
\,
e^{-2\pi i \, jk/l},
\qquad
k = 0,1,\dots,l-1.
\end{equation}

Introducing the unitary Fourier matrix
$F_l \in \mathbb{C}^{l \times l}$ with entries
\begin{equation}
(F_l)_{kj}
=
\frac{1}{\sqrt{l}}
e^{-2\pi i \, jk/l},
\end{equation}
the one-dimensional DFT can be written compactly as
\begin{equation}
\hat{\mathbf{v}} = F_l \mathbf{v}
\qquad \text{with} \qquad
F_l^\dagger F_l = I_l.
\end{equation}

Given a matrix $M \in \mathbb{C}^{m \times n}$, the
two-dimensional DFT is defined by applying the transform
along both dimensions,
\begin{equation}
\hat{M}_{k\ell}
=
\frac{1}{\sqrt{mn}}
\sum_{i=0}^{m-1}
\sum_{j=0}^{n-1}
M_{ij}
\,
e^{-2\pi i
\left(
\frac{ik}{m}
+
\frac{j\ell}{n}
\right)},
\end{equation}
which, by the factorisation of the Fourier kernel, admits
the compact bilinear form
\begin{equation}
    \label{eq:2dfft_matrix}
\hat{M} = F_m \, M \, F_n^\top.
\end{equation}
As described in Secs.~\ref{sec:cooleytukey} and \ref{sec:cv-qfl}, this separable structure is exploited by acting independently on the two registers, so that the row- and column-wise transforms can be performed in parallel.

\subsection{Code implementation}

The encoding circuit is constructed within a Strawberry Fields
\texttt{Program} acting on $m + n$ modes.
The three stages, Two-Mode Squeezing, interferometer on $r_1$,
and interferometer on $r_2$, are implemented as follows:

\begin{lstlisting}[language=Python]
prog = sf.Program(m + n)
with prog.context as q:
    # Stage 1: Two-Mode Squeezing
    for i in range(k):
        S2gate(r[i], 0) | (q[i], q[m + i])

    # Stage 2: Interferometer on register r1
    Interferometer(U) | q[:m]

    # Stage 3: Interferometer on register r2
    Interferometer(Vt.T) | q[m:]
\end{lstlisting}

The quantum circuit realization is illustrated in Fig.~\ref{fig:cv_fourier_layer}. \\
The covariance matrix of the resulting Gaussian state is computed by running the circuit on the Gaussian backend:

\begin{lstlisting}[language=Python]
eng = sf.Engine("gaussian")
state = eng.run(prog).state
sigma = state.cov()
\end{lstlisting}
The encoded matrix is then retrieved from the cross-block:
\begin{lstlisting}[language=Python]
D_enc = sigma[:m, m:m+n]                          # real part
\end{lstlisting}
as described in Eq. \ref{eq:encoded}.
The Gaussian backend does not simulate the full quantum state vector;
instead, it propagates only the first and second moments of the Wigner
function under symplectic transformations.
This results in runtimes that scale polynomially with the number of
modes, making the simulation orders of magnitude faster than a
full Fock-space or wavefunction approach.
The efficiency of the Gaussian backend makes it particularly well-suited
for quantum machine learning applications: the covariance matrix can be
differentiated with respect to the circuit parameters (squeezing
strengths, beamsplitter angles) via standard automatic differentiation
frameworks, enabling gradient-based optimisation of the encoding
without any quantum hardware overhead.

%------------------------------------------------------------------
\paragraph{Cooley-Tukey Code Implementation}\label{app:code}
%------------------------------------------------------------------

The two-dimensional QFT is implemented by appending the CT butterfly circuit to an existing program.
Each butterfly stage applies a phase rotation followed by a beamsplitter:

\begin{lstlisting}[language=Python]
def butterfly_block(self, q, i, j, phase):
    Rgate(phase + np.pi) | q[j]   # phase on odd wire
    BSgate(np.pi/4, 0)  | (q[i], q[j])
\end{lstlisting}

The full 2D QFT is obtained by running independent 1D CT
circuits on the two registers, see Fig. \ref{fig:cv_fourier_layer}:

\begin{lstlisting}[language=Python]
qft = CVQFT()
with prog.context as q:
    qft.qft_1d(q[:m],  register_size=m)   # QFT on r1
    qft.qft_1d(q[m:],  register_size=n)   # QFT on r2
\end{lstlisting}

After the QFT, the two-dimensional Fourier transform of $D$, see Eq. \ref{eq:QFT}, is
read from the updated covariance matrix as:
\begin{lstlisting}[language=Python]
sigma_p  = eng.run(prog).state.cov()
D_hat    = sigma_p[:m, m:m+n] + 1j * sigma_p[:m, (2*m+n):(2*m+2*n)]
\end{lstlisting}
In this way, we recover exactly the 2D QFT we read in Eq. \ref{eq:QFT}.

\bibliography{sn-bibliography}

\end{document}